\documentclass[11pt]{revtex4-1}
\usepackage{amssymb, amsmath, graphicx}

\def\apjs{ApJS}
\def\apj{ApJ}
\def\apjl{ApJ Lett}

\def\aap{A\&A}

\def\mnras{MNRAS}

\def\hunit{\mathrm{\,km\, s^{-1} Mpc^{-1}}}

\begin{document}

\title{Can Non-standard Recombination Resolve the Hubble Tension?}
\author{Miaoxin Liu}
\affiliation{School of Physics and Astronomy, Sun Yat-sen University, 2 Daxue Road, Tangjia, Zhuhai, China}
\author{Zhiqi Huang}
\email{huangzhq25@mail.sysu.edu.cn}
\affiliation{School of Physics and Astronomy, Sun Yat-sen University, 2 Daxue Road, Tangjia, Zhuhai, China}
\author{Xiaolin Luo}
\affiliation{School of Physics and Astronomy, Sun Yat-sen University, 2 Daxue Road, Tangjia, Zhuhai, China}
\author{Haitao Miao}
\affiliation{School of Physics and Astronomy, Sun Yat-sen University, 2 Daxue Road, Tangjia, Zhuhai, China}
\author{Naveen K. Singh}
\affiliation{School of Physics and Astronomy, Sun Yat-sen University, 2 Daxue Road, Tangjia, Zhuhai, China}
\author{Lu Huang}
\affiliation{School of Physics and Astronomy, Sun Yat-sen University, 2 Daxue Road, Tangjia, Zhuhai, China}
\begin{abstract}

The inconsistent Hubble constant values derived from cosmic microwave background (CMB) observations and from local distance-ladder measurements may suggest new physics beyond the standard $\Lambda$CDM paradigm. It has been found in earlier works that, at least phenomenologically, non-standard recombination histories can reduce the $\gtrsim 4\sigma$ Hubble tension to $\sim 2\sigma$. Following this path, we vary physical and phenomenological parameters in RECFAST, the standard code to compute ionization history of the universe, to explore possible physics beyond standard recombination. We find that the CMB constraint on the Hubble constant is sensitive to the Hydrogen ionization energy and  $2s \rightarrow 1s$ two-photon decay rate, both of which are atomic constants, and is insensitive to other details of recombination. Thus, the Hubble tension is very robust against perturbations of recombination history, unless exotic physics modifies the atomic constants during the recombination epoch.
  
\end{abstract}

\maketitle

\section{\label{sec:level1}Introduction}

The tremendous efforts made by cosmic microwave background (CMB) experiments facilitate us to explore the early as well as the current universe. The analysis of the temperature and the polarization anisotropy spectra hints for a spatially flat universe filled with $\sim 5\%$ known matter in the standard chart of particle physics, $\sim 26\%$ cold dark matter (CDM) whose particle-physics nature is yet to be determined, and $\sim 69\%$ dark energy with repulsive gravity, whose microscopic nature is usually interpreted as the cosmological constant $\Lambda$ or equivalently the vacuum energy~\cite{Planck2018Overview}. Within this $\Lambda$CDM paradigm and using the latest CMB data from the Planck satellite, the Planck collaboration was able to precisely determine the cosmological parameters~\cite{Planck2018Params}, among which the current expansion rate of the universe, namely the Hubble constant is constrained to be $H_0 = 67.4 \ \pm 0.5 \hunit$. 

The remarkable success of the $\Lambda$CDM paradigm, however, is challenged by the recent measurements of the local expansion rate of the universe. The local Hubble constant inferred from low-redshift distance ladder (SH0ES), $H_0=74.03\pm 1.42\hunit$, is in $4.4\sigma$ tension with the CMB+$\Lambda$CDM result~\cite{Riess16,Riess18a, Riess18b, Riess19}. More recently, another independent class of $H_0$ measurements using the time-delay of strong-lensing quasar images (H0LiCow, STRIDES) starts to join the high-$H_0$ camp~\cite{H0LiCow, STRIDES2019}. The combination of SH0ES and H0LiCow results gives $H_0=73.8\pm 1.1\hunit$, which raises the low-$z$ (redshift $z\lesssim 1$) and high-$z$ (redshift $z\sim 1100$) Hubble tension to $5.3\sigma$~\cite{H0LiCow}.

The ever increasing Hubble tension, if not due to some unknown observational biases and systematics~\cite{HST70p6, Addison15}, may suggest a fundamental flaw in the standard $\Lambda$CDM paradigm. The possibility of discovering new physics beyond $\Lambda$CDM has inspired discussion of many theoretical models, most of which are related to the dark components~\cite{Miao_2018, Lin_2018, Lin_2019, Sola_2019,Rossi:2019lgt, Karwal_2016,Alexander_2019,Poulin_2019, DV_2017,Yang_2018a,Yang_2018b,DV_2018,Bhattacharyya_2019, DV_2019, Agrawal_2019_Rock, Agrawal_2019_H0}. Other possibilities are also intensively investigated~\cite{D_Eramo_2018, Benetti_2018,Graef_2019,Carneiro_2019, Zhang19, Bolejko_2018,SuperCMB}. Ref.~\cite{LH19} studied the impact of non-standard primordial fluctuations from inflation, and found that the CMB $H_0$ constraint is insensitive to the primordial conditions, thus excluding early-universe origin of the Hubble tension for a broad class of models. Following a series of attempts to explore non-standard recombination~\cite{Farhang12,Farhang13,Planck2015Params}, Ref.~\cite{Chiang18} found that phenomenological modification to the timing and width of the recombination process can significantly reduce the Hubble tension to $\sim 2\sigma$. However, unlike the primordial fluctuations that are related to unknown physics at very high energy scales $\gtrsim 10^{14}\mathrm{GeV}$, the recombination process depends on well-tested physics at energy scales $\lesssim \mathrm{eV}$. The phenomenological perturbations to the ionization fraction function $X_e(z)$ in Ref.~\cite{Chiang18} as well as in earlier works~\cite{Farhang12,Farhang13,Planck2015Params} may be too ``nonphysical'' to comply with the basic physical pictures of recombination. To tackle this problem, in this paper we use a very different approach to perturb recombination. We vary the parameters in the system of equations that govern the evolution of $X_e(z)$. This method automatically retains the major physical structure of Helium and Hydrogen recombination processes, and directly connects the perturbations in $X_e(z)$ to underlying physical parameters. We can then identify the physical degrees of freedom that may relieve the $H_0$ tension.
\section{Method \label{sec:method}}

The standard RECFAST code based on a series of work~\cite{Peebles68, Seager99} contains a fast and approximate algorithm to compute the helium and hydrogen recombination processes. Despite the presence of a few phenomenological ``fudge factors'' and a few fitting parameters in RECFAST, it is believed that sub-percent level accuracy can be achieved after careful calibration against more detailed calculations~\cite{CosmoREC, Seager00, Wong08, Scott_2009}.

In Table~\ref{tab:params} we list the major (non-cosmological) parameters involved in RECFAST algorithm. 
\begin{table}
  \caption{Parameters in RECFAST \label{tab:params}}
  \begin{tabular}{lll}
    \hline
    \hline
    parameter & definition & precision \\
    \hline 
    $E_{\rm H,  energy}$ & Hydrogen ionization energy & theoretically and experimentally determined constant \\
    $A_{\rm H, 2\gamma}$ & Hydrogen $2s\rightarrow 1s$ two-photon rate & theoretically determined constant \\
    $f_H$ & Hydrogen fudge factor & phenomenological, has some uncertainties \\    
    $A_{\rm He, 2\gamma}$ & Helium $2s\rightarrow 1s$ two-photon rate & theoretically determined constant \\    
    $f_{\rm He}$ & Helium fudge factor & phenomenological, has some uncertainties \\
    \hline
  \end{tabular}
\end{table}
In principle the HeI and HeII ionization energies should also be included in the list. However, the prediction of CMB observable is not sensitive to the helium parameters. Including too many helium parameters would be redundant for CMB analysis. In the latest version RECFAST V1.5.2, there are many more parameters to capture details of Helium recombination. These parameters only lead to tiny ($\ll$ percent) corrections to CMB observables and have little impact on cosmological parameters. Thus, hereafter we will only discuss the parameters listed in Table~\ref{tab:params}.

If we admit the precision and robustness of RECFAST code in the $\Lambda$CDM paradigm, none of the parameters in Table~\ref{tab:params} has much room to vary. Physics beyond $\Lambda$CDM may lead to re-calibration of the phenomenological fudge factors $f_H$ and $f_{\rm He}$, or even make them time-dependent. The variation of atomic constants $E_{\rm H,  ion}$, $A_{\rm H, 2\gamma}$ etc., however, would be much more difficult to achieve and is considered to be rather ``controversial''. To fully explore the impact of non-standard recombination on $H_0$, we nevertheless relax the atomic constants in part of our analysis, too. More concretely, we rescale the parameters in Table~\ref{tab:params} with scaling factors $s_{\rm H, energy}$, $s_{\rm H, 2\gamma}$, $s_{\rm H, fudge}+n_{\rm H, fudge}(a-a_\ast)$, $s_{\rm He, 2\gamma}$, $s_{\rm He, fudge}$, respectively, where $a$ is the cosmological scale factor normalized to unity today, and $a_\ast = 1/1090$ is its value at recombination in standard scenario.  We are allowing some time-dependence of the hydrogen fudge factor, because it governs the evolution of $X_e$ for a much longer time (till the late universe). In total we have six recombination parameters. We assume a uniform prior in $[0.8, 1.2]$ for each scaling  parameter $s_{\ldots}$, and uniform prior in $[-0.2, 0.2]$ for the running parameter $n_{\rm H, fudge}$. These parameters are added to the standard CosmoMC~\cite{CosmoMC} for a full MCMC analysis with all the parameters varying. We use the Planck final release TTTEEE + lowE + lensing likelihood ~\cite{Planck2018Like} and apply a lower-bound to the reionization redshift: $z_{\rm reion} > 6$, which is independently inferred from Gunn–Peterson trough~\cite{Becker_2001}. The reionization prior can effectively control the degeneracy between reionization and the non-standard RECFAST $X_e$.

We modify the publicly available CosmoMC package to add the aforementioned six extra recombination parameters to the standard six cosmological parameters: the baryon density $\Omega_bh^2$, the CDM density $\Omega_ch^2$, the angular extension of sound horizon on the last scattering surface $\theta$, the reionization optical depth $\tau$, the amplitude $A_s$ and tilt $n_s$ of the primordial scalar power spectrum. We do a MCMC run with all the parameters varying, as well as a ``less controversial'' run with the fudge factors varying but with the atomic constants fixed ($s_{\rm H, energy}=s_{\rm H, 2\gamma}=s_{\rm He, 2\gamma}=1$). We dub the two runs ``vary-all'' and ``vary-fudge'', respectively.

\section{Results \label{sec:results}}

\begin{figure*}
\centering
\includegraphics[width=1\textwidth]{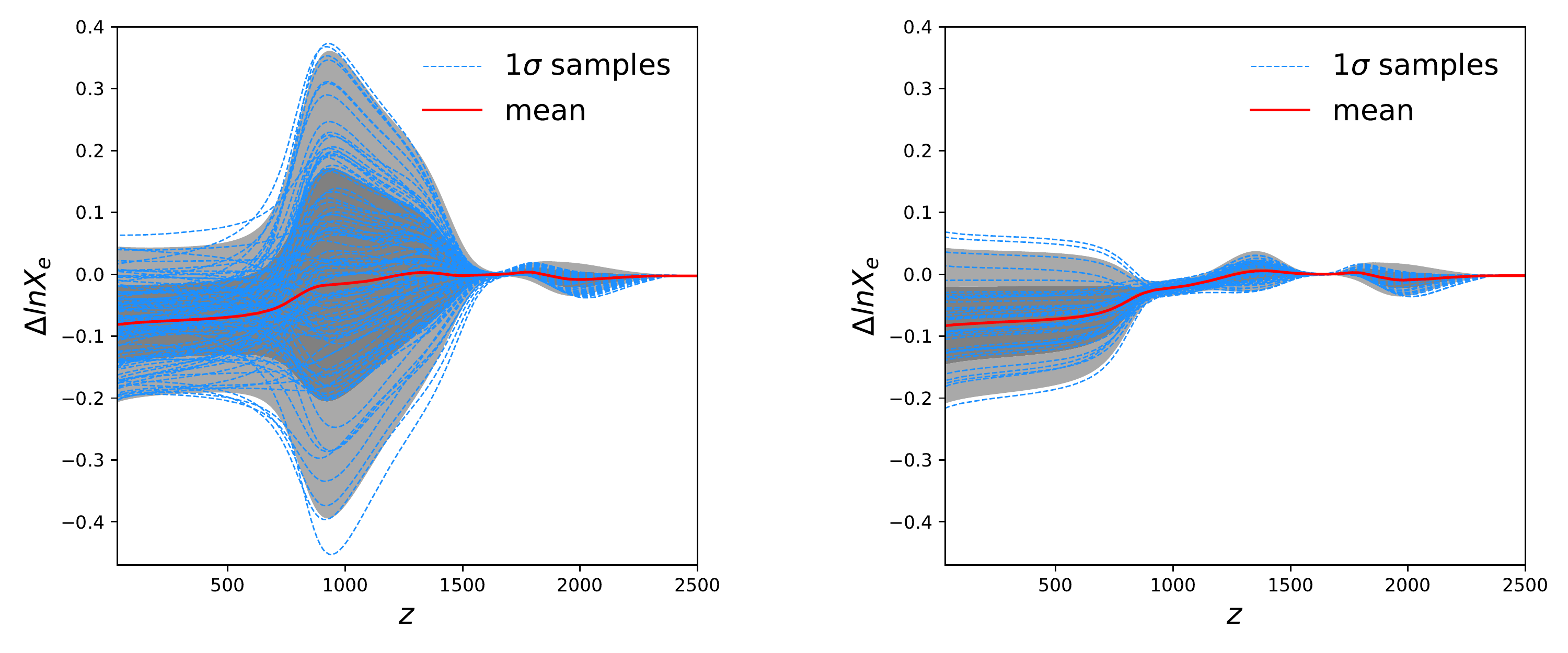}
\caption{Left panel: trajectories of $\ln X_e(z)$ randomly selected from the top 68.3\% likelihood-ranked MCMC samples from the ``vary-all'' run. For better visibility we have subtracted $\ln X_e(z)$ of the reference model - Planck best-fit $\Lambda$CDM~\cite{Planck2018Params}. The dark- and light-gray areas are marginalized 68.3\% and 95.4\% confidence-level regions, respectively. Right panel: the same as the left panel, except that only samples with restricted hydrogen recombination redshift $1089.156<z_\star<1090.335$ (Planck + $\Lambda$CDM $\sim 3\sigma$ bounds) are used. \label{fig:scat}}
\end{figure*}

We take the Planck best-fit $\Lambda$CDM~\cite{Planck2018Params} as a reference model, and show the variations of $\ln X_e$ trajectories in the left panel of Figure~\ref{fig:scat}. Unlike the parametrization used in Ref.~\cite{Chiang18} where the $X_e$ variations are mainly due to the shift of hydrogen recombination redshift $z_\star$, defined by $X_e(z_\star)= 0.5$, our parametrization explores many more degrees of freedom. To more explicitly demonstrate this,  in the right panel of Figure~\ref{fig:scat} we show, by selecting sub-samples with $z_\star$ almost frozen, the rich structures in $\ln X_e(z)$ due to the degrees of freedom beyond a $z_\star$ shift. 

\begin{figure*}
\centering
\includegraphics[width=0.48\textwidth]{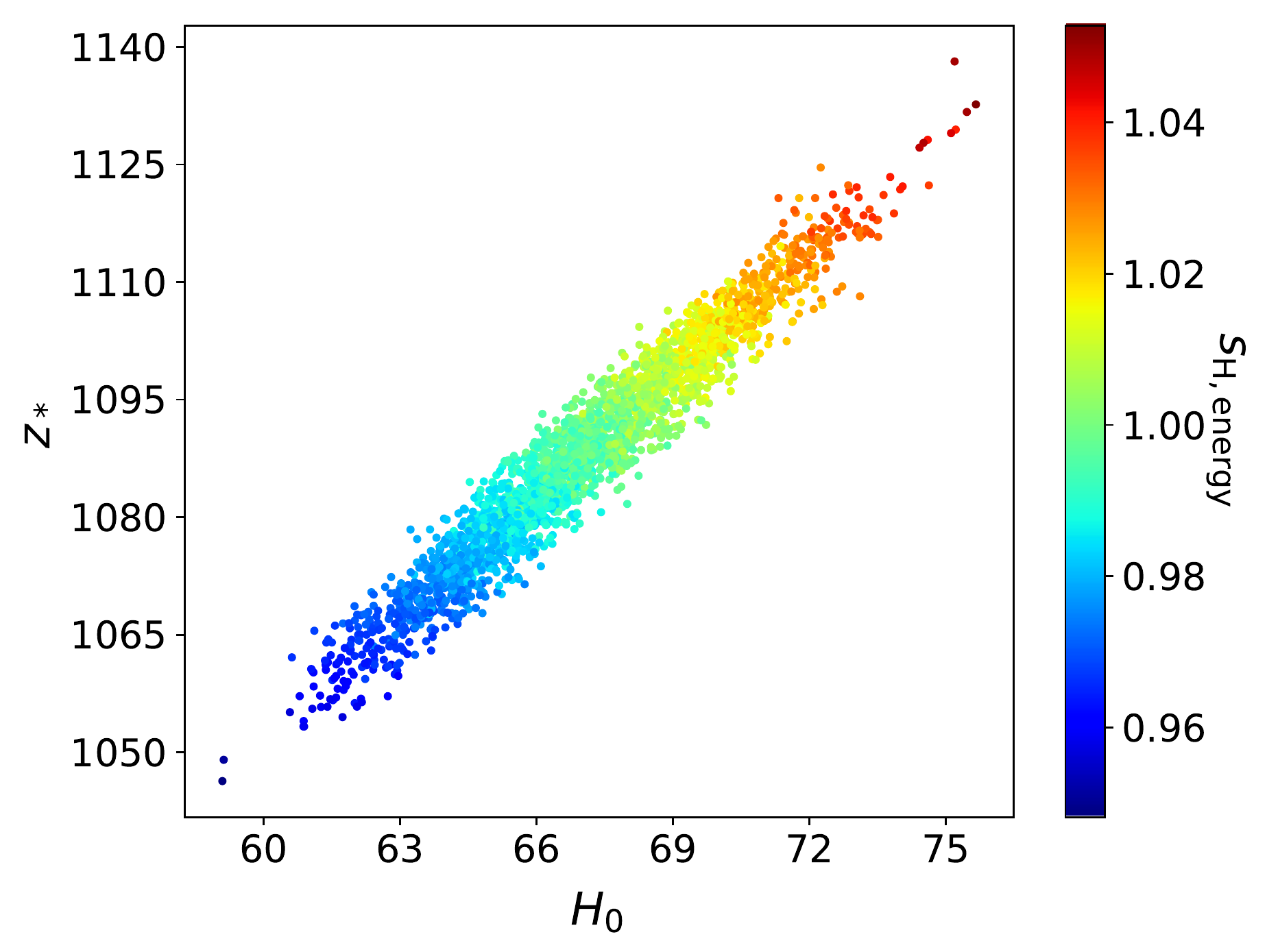}\includegraphics[width=0.48\textwidth]{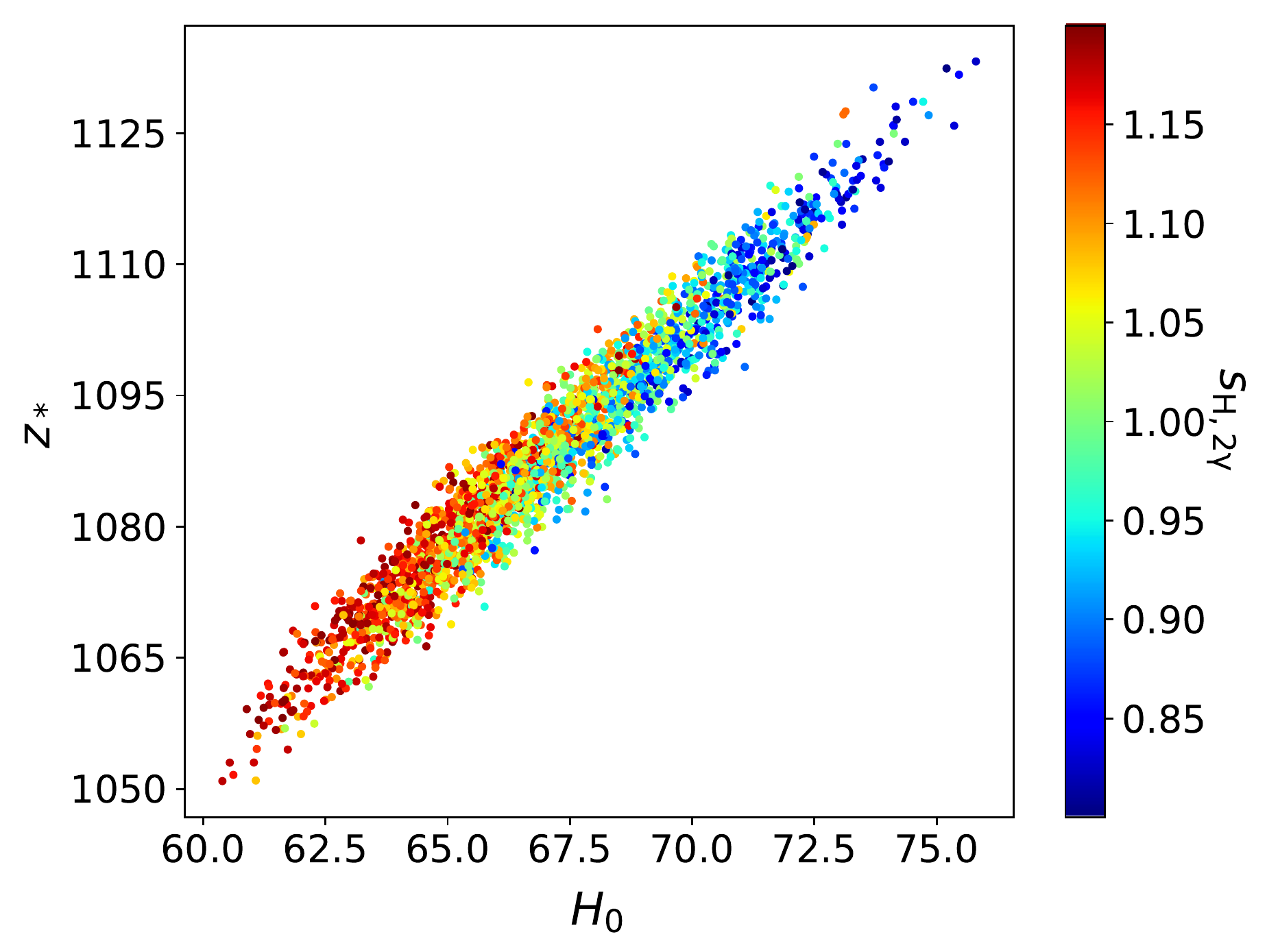}
\caption{The marginalized posterior distribution of $z_\star$ and $H_0$ for the ``vary-all'' run. The left and right panels show the dependence of the distribution on the hydrogen ionization energy parameter  $s_{\rm H,energy}$ and on the hydrogen $2s\rightarrow 1s$ two-photon rate parameter $s_{H,2\gamma}$, respectively. \label{fig:zstar}}
\end{figure*}

The hydrogen recombination redshift $z_\star$ is indeed a key quantity connected to the Hubble tension. In Figure~\ref{fig:zstar} we demonstrate the strong degeneracy between $z_\star$ and $H_0$ for the ``vary-all'' run. We find that the Hubble tension can be significantly relieved when $z_\star$ is set free, in agreement with Ref.~\cite{Chiang18}. However, we find that significant relaxation of $z_\star$ can only be achieved by variation of the hydrogen ionization energy or the hydrogen $2s\rightarrow 1s$ two-photon decay rate, both of which are well-determined atomic constants by quantum mechanics.

\begin{figure*}
\centering
\includegraphics[width=0.5\textwidth]{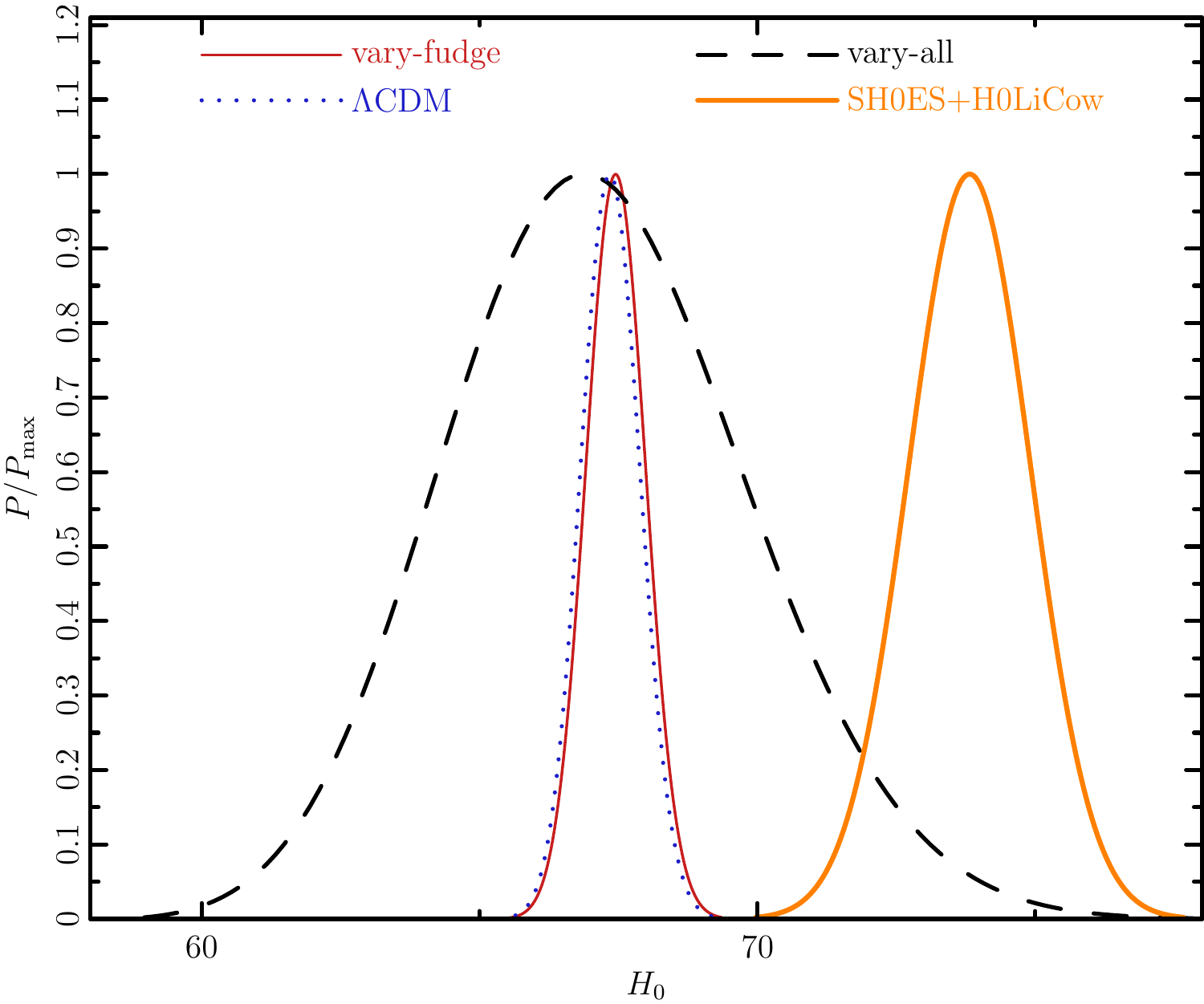}
\caption{The marginalized posterior distribution of $H_0$. \label{fig:H0}}
\end{figure*}

The final marginalized constraints on $H_0$ are shown in figure~\ref{fig:H0}, for both ``vary-all'' and ``vary-fudge'' runs. The relaxation of atomic constants in ``vary-all'' run frees the hydrogen recombination redshift and hence significantly worsens the constraints on $H_0$. This can be understood because the CMB $H_0$ constraint is mainly determined by the angular diameter distance to the last scattering surface, i.e., the $z=z_\star$ surface. In the less controversial ``vary-fudge'' run with the atomic constants frozen, the constraint on $H_0$ does not significantly differ from the standard $\Lambda$CDM. The Hubble tension with SH0ES + H0LiCow measurement persists at a $5.2\sigma$ level.

\section{Conclusion \label{sec:conclusion}}

By connecting the perturbations in ionization history to physical parameters, we found that exotic variations of atomic constants must be claimed in order to have a significant relaxation of the CMB constraint on the Hubble constant. If one admits the  basic physical properties of hydrogen, there is no obvious room to vary the recombination history to relieve the Hubble tension. Our result does not contradict with previous findings in Ref.~\cite{Chiang18}. Rather, we point out how ``non-standard'' the recombination process has to be in order to serve as a buffer belt between the low- and high-redshift Hubble drivers.

%\bibliographystyle{aasjournal}
%\bibliography{cites}

\end{document}